# The senses considered as one perceptual system

Thomas A. Stoffregen[1], Bruno Mantel[2], Benoît G. Bardy[3]

[1]University of Minnesota, Minneapolis, MN, USA
[2]CesamS, Normandie Université, Université de Caen Normandie, France
[3]EuroMov, Université de Montpellier, Institut Universitaire de France

# The Senses Considered as One Perceptual System

Thomas A. Stoffregen[1], Bruno Mantel[2], Benoît G. Bardy[3]

[1]School of Kinesiology, University of Minnesota;
[2]CesamS, Université de Caen Normandie;
[3]EuroMov, Université de Montpellier, Institut Universitaire de France

**Abstract**
J. J. Gibson (1966) rejected many classical assumptions about perception, but retained one that dates back to classical antiquity: The assumption of separate senses. We suggest that Gibson's retention of this assumption compromised his novel concept of perceptual systems. We argue that lawful, 1:1 specification of the animal-environment interaction, which is necessary for perception to be direct, cannot exist in individual forms of ambient energy, such as light, or sound. We argue that specification exists exclusively in emergent, higher order patterns that extend across different forms of ambient energy. These emergent, higher order patterns constitute the global array. If specification exists exclusively in the global array, then direct perception cannot be based upon detection of patterns that are confined to individual forms of ambient energy and, therefore, Gibson's argument for the existence of several distinct perceptual systems cannot be correct. We argue that the senses function as a single, irreducible perceptual system that is sensitive exclusively to patterns in the global array. That is, rather than distinct perceptual systems there exists only one perceptual system.

*The Senses Considered as Perceptual Systems* (J. J. Gibson, 1966) remains a landmark achievement. In it, Gibson rejected major portions of classical theory on perception, offering in their place qualitatively new arguments. Yet, Gibson's rejection of classical assumptions was not complete. In this article, our focus is on relations between the senses. We argue that the 1966 book contains two different theories about relations between the senses. At some places in the book, Gibson advocated a view of intersensory relations that was similar to arguments offered by Aristotle. However, at other places in the book, Gibson offered a different interpretation, one that is incompatible with the Aristotelian view. We claim that these two views are incompatible, that is, we argue that Gibson made assertions that are mutually exclusive. We argue that Gibson's Aristotelian view should be rejected, and that it is incompatible with the concepts of specification and direct perception. In addition, we claim that, taken to its logical conclusion, Gibson's novel, non-Aristotelian argument mandates a broader, more general, more unitary and, ultimately, more successful view of perception, in which the senses operate as a single, irreducible, perceptual system.

Corresponding author:  Thomas A. Stoffregen   tas@umn.edu
School of Kinesiology, Cooke Hall 224, 1900 University Ave. SE, Minneapolis MN 55455

Our argument is an elaboration of a theory presented by Stoffregen and Bardy (2001). That article was written for a general audience, whereas in the present article we focus on relations between perception and the Ecological Approach to Perception and Action. Like Gibson (1966), our argument is founded upon a consideration of ecological physics; the nature of the animal-environment interaction, and how it is specified. Ambient arrays are structured by the *animal-environment interaction* (that is, by the position and motion of the animal relative to its environment), and this structuring is governed by physical laws (i.e., laws of the propagation, reflection, and absorption of energy) in such a way that aspects of the animal-environment system give rise to unique structures or patterns in ambient energy. This leads to the hypothesis that these patterns in ambient energy are sufficient for accurate perception because the animal-environment interaction is *specified* in the spatiotemporal structure of ambient arrays. Specification refers to a lawful, 1:1 relation between patterns in ambient arrays and the aspects of the animal-environment interaction that give rise to them (Shaw, Turvey, & Mace, 1982).

We lean heavily on Gibson's (1966) argument that perception should be defined in terms of function, that is, in terms of the aspects of reality that the perceiver should seek to detect. Following Gibson, we argue that behavior is perceived and controlled relative to real aspects of the animal-environment system, rather than being controlled relative to sensory reference frames, internal models, estimates, or representations. The referents for perception and control are outside the head, in the animal-environment system (cf., Mace, 1977). We argue that the relevant ecological physics always comprise the activity of the organism relative to multiple independent referents and that this activity structures different forms of ambient energy. For this reason, we argue that behavior always requires control relative to referents that cannot be specified in patterns that exist within any single type of ambient energy. We argue that perceivers should always seek information about behavior relative to multiple, independent referents. We argue that the needed information exists, but that it exists solely in emergent patterns that extend across multiple forms of ambient energy, such that accurate, direct perception must always be based upon sensitivity to these emergent, higher order patterns. We claim that perception naturally functions to obtain information about these emergent, higher-order patterns extending across multiple forms of ambient energy. We conclude that, in one respect, Gibson (1966) was mistaken: Perception does not comprise multiple, overlapping systems. Rather, we claim, perception consists of a single, irreducible perceptual system.

Above all, Gibson (1966), understood that his presentation was a beginning rather than an end; the first statement of a new theory that could be completed and rendered fully coherent and consistent only through future development: "The answers to these questions are not yet clear, but I am suggesting new directions in which we may look for them" (p. 5). In this article, our primary aim is an effort toward greater clarity on some of Gibson's questions.

## *The Senses Considered*: Selective summary

*The Senses Considered as Perceptual Systems* was novel in many ways. In this section, we briefly review portions of Gibson's argument that are relevant to our analysis. Gibson elected to begin his book (Chapter I) with a discussion of the contents of perception, that is, the world that we perceive. For a psychologist to offer an analysis of the environment was without precedent, a fact that contemporary readers understood. For example, von Fieandt (1967, p. 230) acknowledged, "no material like the contents of Chapters I-III has ever appeared before in textbooks of psychology". For a book about perception to *begin* with such an analysis was even more exceptional (Lombardo, 1987; Reed, 1988; Sherrick, 1967). Analysis of the world to be perceived, that is, of the real, physical world at the scale of the animal-environment system, came to be known as *ecological physics*.

A second important aspect of Gibson's argument was his claim that perception is active. For centuries, scholars had assumed that perception was the response of sense receptors to energy stimuli imposed upon them. Gibson acknowledged that stimuli could be imposed on passive observers, but he pointed out that this situation was simply unrepresentative. In ordinary life, perceivers actively seek out information. In particular, we select the patterns of energy that stimulate our sense receptors. Accordingly, Gibson formulated a theory of perception that was based upon stimulation that was obtained rather than being imposed, upon the active pickup of information rather than upon passive responses to stimuli. This was a qualitative shift in the nature of perception (in the idea of what perception actually was), a break with tradition so complete that many contemporary readers found it to be incoherent (Reed, 1988).

A third important aspect of Gibson's argument was his claim that perception is based upon ambient information. In *The Senses Considered as Perceptual Systems*, Gibson (1966) provided the first general argument about the concept of ambient information, and the concept of specification. He abandoned the traditional assumption that variations in sensory stimuli bear a relation to physical reality that is intractably ambiguous and he proposed, instead, that these variations are related to facts of physical reality through natural law, such that they provide veridical information about reality. This argument was the basis for Gibson's claim that perception can be direct.

Throughout the book, Gibson (1966) offered numerous descriptions of particular types of ambient information. A consistent feature of his examples was that the information in question resided not in lower order elements of stimulation but, rather, in higher order relations, that is, in emergent patterns of stimulation. From this book, perhaps the best-known example is global optic flow that arises from motion of the point of observation relative to the illuminated environment. Gibson pointed out that the information about this "self-motion" does not exist in any of the optical velocity vectors arising from motion relative to individual texture elements in the environment. Rather, the information exists solely in patterns of relative velocity. In other words, the information is an emergent property of relations between individual velocity vectors. A familiar analogy is the figure of a triangle, which does not exist in any of a trio of lines, but emerges (literally, comes into existence) as a function of higher order relations between the lines. This idea, that the whole (i.e., global optic flow, or a triangle) differs qualitatively from the sum of the parts (i.e., individual velocity vectors, or a trio of lines) was central to Gibson's concept of specification and, therefore, to his claim that perception can be direct.

Finally, in 1966, Gibson introduced the concept of a perceptual system, as opposed to the classical concept of a sense. The classical concept referred to sensory receptors, such as the retina, and the cochlea, and to the neural structures that connect those receptors to the brain. In this conception, perception must consist of neural responses to stimuli that impinge upon receptors. Movement, or other acts of the organism were understood to be subsequent to perception. By contrast, in Gibson's concept of a perceptual system perception is an act, a controlled seeking out of information: The organism uses its own activity to orient receptors toward things of interest in the animal-environment interaction, to seek out information that is relevant to the organism's behavioral goals. For this reason, motor behavior (and motor effectors) is part of the perceptual system.

## Critique of the *Senses Considered*

Living things are surrounded by a sea of energy. This energy is not homogeneous; rather, it exists in several forms. There are forces, such as gravity, and inertial force which, at the ecological scale combine to yield a single (though changing) *gravitoinertial force*. There is electromagnetic energy, extending across a spectrum that includes infrared, light, radio waves, x-rays, and so on. There is chemical energy in aerosols and solutes. Interactions between gravitoinertial force and substances (e.g., solids, liquids, or gases) yield mechanical energy, which spans a spectrum that extends from pushing and pulling forces all the way to ultrasound. These various forms of energy are ambient for all living things, at all times.[1] In this context, any movement of an organism will alter the structure of multiple forms of ambient energy. The resulting changes comprise the dynamic flux of energy to which the behaving organism is exposed. Gibson asserted that this dynamic flux contains veridical information about the animal-environment interaction, and that the pickup of this information enables perception that is both accurate and direct. It was in this context that he presented the theory that perception happens through perceptual systems. Our concern is with Gibson's choice of the plural, *systems*, rather than the singular, *system*. Within the Ecological Approach to Perception and Action, what motivates the assumption that there exist distinct perceptual systems, rather than a single, unitary perceptual system?

In many ways, Gibson chose to break from received views, repeatedly rejecting assertions that had been so uncontroversial as to appear self-evident. He pointed out that these assertions, often, were assumptions, that is, claims that had not been subjected to direct testing, either empirical or analytic. Yet Gibson himself was not immune to this temptation. He accepted one of the most ancient, most pervasive of assumptions about perception, one that had been subjected to minimal critical evaluation.

### *Gibson and Aristotle*

*The Senses Considers as Perceptual Systems* famously begins with a quotation from Aristotle, on a page that ends with Austin's disparagement of "assertions that are not true" (Gibson, 1966). Among other things, this page makes clear that Gibson traced important aspects of perceptual theory to Aristotle, that he was willing to question these Aristotelian assumptions, and to reject ones that could not stand up to scrutiny.

However, Gibson (1966) did not explicitly reject the entire Aristotelian program. One claim that Gibson did not reject was Aristotle's distinction between *special objects* of perception and *common sensibles*. According to Aristotle, each sense has its special objects, "that which cannot be perceived by any other sense than that one in respect of which no error is possible; in this sense color is the special object of sight, sound of hearing, flavor of taste," (1931, p. 418a). The special objects of perception contrast with common sensibles, which are "perceptible by any and all of the senses" (1931, p. 418a). Among the common sensibles are movement, rest, number, figure, and magnitude. Gibson rejected Aristotle's hypothesis that there exist different modes of sensation. However, in elaborating his concept of perceptual systems he adopted a distinction between information available uniquely to some perceptual systems and percepts available redundantly or equivalently to different perceptual systems:

> Different stimulus energies—acoustical, chemical, radiant—can all carry the same stimulus information . . . patterns in the flux of sound, touch, and light from the environment may be equivalent to one another by invariant laws of nature (Gibson, 1966, p. 55).

> The material *color* (pigmentation) of a surface is not tangible but only visible. The relative temperature, however, is tangible but *not* visible (italics in the original, Gibson, 1966, p. 123).

> The only trustworthy information for locomotion in the world is *visual* kinesthesis. Thus, it is only by observing on what part of the world the focus of optical expansion falls that one can see where he is going (italics in the original,
> Gibson, 1966, p. 200; cf. E. J. Gibson, 1983; Lombardo, 1987, p. 38).

---

[1] Exceptions are surprisingly rare. Examples include freefall or weightlessness, where there may be no mechanical energy, and caves or mines, which may exclude most electromagnetic energy (this is why deep mines are chosen as venues for neutrino detectors; e.g., Sanchez et al., 2003).

## *The assumption of separate senses*

Aristotle (1931, p. 425b) asked "why we have more senses than one". He argued that multiple senses exist "to prevent a failure to apprehend the common sensibles . . . The fact that the common sensibles are given in the objects of more than one sense reveals their distinction from each and all of the special sensibles" (p. 425b). That is, Aristotle argued that multiple, distinct senses must exist so that we may distinguish percepts that are redundant across the senses from percepts that are unique to a particular sense.

This *assumption of separate senses* is, we argue, just that; an assumption. By asking *why* we have more senses than one, Aristotle made the implicit assumption that there *are* more senses than one. Discussion of how (or whether) senses (or perceptual systems) overlap is coherent only under the prior assumption that there *are* multiple senses or perceptual systems, that is, that perception exists as a set of distinct senses or systems, and not as a single, unitary system. Similarly, Aristotle's distinction between special objects and common sensibles is coherent if and only if we first assume that separate senses actually exist. For the same reason, Gibson's argument that the perceptual systems are integrated and cooperative entails the Aristotelian assumption that separate perceptual systems (plural) actually exist: Without separateness, there is nothing to integrate, no distinct systems that might (or might not) cooperate. While Gibson rejected Aristotle's ideas about the nature of the senses, he accepted the Aristotelian assumption of their plurality; that is, Gibson accepted the Aristotelian assumption of separate senses. Like Aristotle, Gibson did not present an explicit argument for the existence of distinct senses (or perceptual systems, as he conceptualized them). Rather, for Gibson, as for Aristotle, the existence of distinct perceptual modalities was taken for granted.

The origins, history, and pervasiveness of the assumption of separate senses, as well as its possible bases, were discussed at length by Stoffregen and Bardy (2001, Section 2).

## *The Aristotelian legacy, with one exception*

In the 1966 book, the position taken most often is consistent with the Aristotelian distinction between special objects and common sensible (Stoffregen & Bardy, 2001, Section 3.3.2). In many places, Gibson offered characterizations of information available to individual perceptual systems (i.e., special objects of perception; e.g., Chapters V, X). In other places, he characterized information that, he claimed, was available redundantly to different perceptual systems (i.e., common sensibles; e.g., Chapter III).

While Gibson most commonly offered arguments consistent with *special objects* and *common sensibles*, he also articulated another argument that, we claim, differs qualitatively from the Aristotelian position. The most developed instance of this argument appears in Chapter IV, in which he claimed that some types of information existed exclusively as emergent, higher order relations between patterns available to the vestibular and somatosensory systems. In a specific example (p. 62-63), Gibson noted that for an animal resting on a sloping ground, the downward direction of gravity (influencing stimulation of the vestibular system) would be at an angle relative to the direction of resistive forces stimulating the somatosensory system (Figure 1). He referred to this angle as a discrepancy between stimulation of the vestibular and somatosensory systems, and stated, "the discrepancy is information…if it can be registered, the organism will be able to detect the slope of the ground" (p. 63). To assist in illustrating the concept of emergent, higher order relations, in Figure 2 we reformulate Figure 1 in purely geometrical terms. The idea that information exists in higher-order relations between stimulation of different systems is not compatible with the Aristotelian concepts of common sensibles, or special objects (Stoffregen & Bardy, 2001).

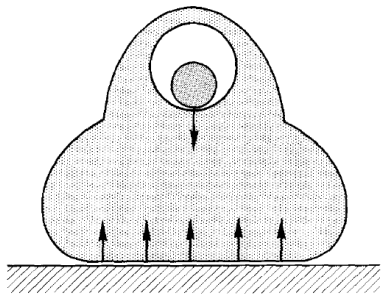
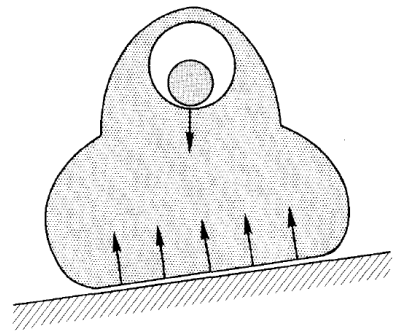

Figure 4.2 The statocyst of an animal resting on a substratum. The pull of gravity and the push of the surface of support are represented by arrows. If the ground is horizontal, the directions coincide.

Figure 4.3 The statocyst of an animal resting on a sloping substratum. The directions of the pull of gravity and the push of the surface of support do not coincide.

**Figure 1.** Information about the slope of the ground, exists in higher-order relations between patterns available to the vestibular system and the somatosensory system. Left: The emergent relation between vestibular and somatosensory stimulation specifies that the ground is perpendicular to the gravitoinertial force vector. Right: The emergent relation between vestibular and somatosensory stimulation specifies that the ground is tilted relative to the gravitoinertial force vector. Reproduced from Gibson (1966).

To summarize, Gibson (1966) made two distinct assertions, but made them in separate parts of the book, and in relation to his discussion of separate issues. At no point in the book did Gibson consider these two claims together. In particular, he offered no discussion of how these two claims might, or could relate to each other. In several places, he suggested that the perceptual systems are "interrelated rather than mutually exclusive" (e.g., p. 47), but he did not reject the Aristotelian concept of common sensibles, and he did not claim that sensitivity to emergent relations extending across multiple forms of ambient energy was a general principle. In short, taken as a whole, the 1966 book is ambiguous with regard to the plurality of the senses. Gibson's concept of a perceptual system, as opposed to sensory systems, is clear and consistent, but the book is ambiguous with respect to the plurality of perceptual systems: Can we (or must we) accept the assumption of separate senses? Are there, in fact multiple perceptual systems? If so, on what basis can we distinguish between them, and how do they relate to one another? And finally, if there is a plurality of perceptual systems, then what is the nature of the ambient information to which they are sensitive? Do patterns in individual forms of ambient energy bear a lawful, unique, 1:1 relation to aspects of reality? In the remainder of this article, we argue that the assumption of separate senses can be and, indeed, must be rejected.

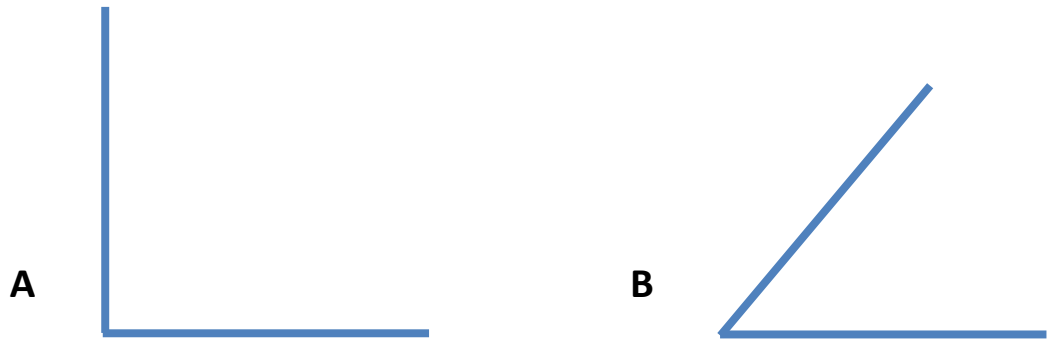

**Figure 2.** Examples of emergent properties in geometry. A. A right angle. B. An acute angle. Angles do not exist in individual lines; rather, they come into existence, or emerge from relations between lines.

## Toward a unitary theory of the senses

In this section, we attempt to develop a view of perception that does not depend upon the assumption of separate senses, or upon concepts consistent with the Arisotelian ideas of *common sensibles* and *special objects* of perception. Like Gibson (1966), we begin with the contents of perception, that is, with what it is that perception is supposed to tell us *about*. We argue that the contents of perception can help us to understand the nature of the information that supports perception which, in turn, can help us to understand how perception works, in general, and how the senses relate to one another, in particular.

### *The animal-environment interaction to be perceived*

In seeking to escape from rejected assumptions, Gibson (1966) found it useful to begin by examining the objects of perception. If we accept that perception yields knowledge (i.e., that perception is epistemological), then we can ask what type of knowledge is obtained. Following Darwinian principles (Lombardo, 1987; Reed, 1988), Gibson reasoned that perception yields (and should yield) knowledge about the environment. For this reason, in writing *The Senses Considered as Perceptual Systems*, Gibson made an organizational decision that was as fateful as it was exceptional: He began the book with a description of the environment. In Chapter I, Gibson considered the environment as a source of stimulation. In his final book, he revised this to a focus on the environment as such, that is, the environment to be perceived (1979/1986). In both 1966 and 1979/1986, his treatment combined aspects of the animal-environment interaction (e.g., locomotion) with aspects of the environment, as such (e.g., substances, surfaces). Our focus is on the animal-environment interaction as the contents of perception (cf. Lombardo, 1987).

Behaviors exist as unitary events. On this point, Eleanor Gibson (1983, p. 19-20) was quite explicit: "Perhaps the most remarkable thing about perception is the unified quality of everything that we perceive… Whence comes the wholeness that is so important in our perceptions of the world?" When a person speaks, it is meaningful and coherent to describe the situation as one in which only one thing is happening—the act of speaking. When we converse, generally we do not care about what the speech sounds like, or looks like; we care about the unitary entirety of what is being communicated. Similarly, when a person walks it is meaningful and coherent to say that only one thing is happening—the act of walking. Typically, the optical and gravitoinertial consequences of walking are of secondary interest; the primary interest is the nature of the resulting translation, such as its speed, direction, and destination. Outside the laboratory, the unitary nature of these events is reflected in our experience of them; we experience speech as such (i.e., we do not experience "visible speech" or "audible speech"), and walking as such (i.e., we do not experience "optical walking", or "gravitoinertial walking"). Each event, speaking and walking, structures multiple forms of ambient energy, and for this reason it is possible to claim that the relevant information exists in energy-specific components, such as patterns of sound, patterns of light, or patterns of force. Yet at the same time each event exists as a single, unitary entity, and it is these entities that are of primary concern to perceivers. The unitary events are logically prior to their energy-specific consequences in the sense that potential sensory stimulation depends on the events for its existence, but the events do not depend upon potential sensory stimulation for their existence.[2] In seeking to understand perception (and to understand relations between the senses) we begin with the assumption that the objects of perception are aspects of the animal-environment interaction. Thus, when we pose questions about relations between the senses the questions are of the form, "How do the senses function to provide information about the animal-environment interaction?"

## *Perception and evolution*

It is important to emphasize that our discussion of perception is general, that is, it is intended to apply across all species (Stoffregen & Bardy, 2001, p. 246). In this context, it is useful to consider the evolution of perception. Recent estimates suggest that life emerged on Earth approximately 3.7 billion years ago (e.g., Fedo & Whitehouse, 2002). On this time scale, the appearance of anatomically distinct receptors is relatively recent. For example, eyes as distinct anatomical structures emerged only about 500 million years ago (Lamb, 2011). Perception-action coupling exists in species that do not have anatomically differentiated receptor systems, such as single celled animals (e.g., Pittenger & Dent, 1988). Given the success of these species and their primacy in evolutionary chronology, to argue that perception comprises separate perceptual systems would appear to require the argument that through evolution living things abandoned the original unity of perception; that early organisms had a single, unitary perceptual system and that evolutionary processes led to the fracturing of this unitary system into the distinct perceptual systems described by Gibson (1966). From an Ecological perspective, how could such an argument be motivated? We suggest that evolution did not abandon the original unity of perception; rather, we suggest that the differentiation of perceptual anatomy emerged in the service of action differentiation, but without a qualitative shift from unity to the existence of distinct perceptual systems. To put it plainly, we claim that perception in large, anatomically differentiated species does not differ qualitatively from perception in smaller species that do not have anatomically differentiated sensory receptors. We are not aware of any treatment of the Ecological Approach to Perception and Action that has addressed this issue; that is, we are not aware of any arguments attempting to explain (or predict) how the unitary perceptual sensitivity of single-celled organisms could (or, in a Darwinian system, should) abandon a successful unitary function and evolve, *de novo*, separate functions. As pointed out by Stoffregen and Bardy (2001, Section 2.3), the existence of anatomically distinct receptor systems does not mandate such an argument.

---
[2] This conception may be similar to Fowler's claim that "distal things are amodal" (2004, p. 197). See also Principle I from Richardson, Shockley, Fajen, Riley, and Turvey (2008).

Eleanor Gibson (1983) argued strongly for the same position in the context of phylogeny, that is, in the development of the individual. Referring to human development, she argued, "the perceptual systems work together from the start" (p. 25), and that "modal properties, like color, are … differentiated or dissociated from the whole, which is more primitive" (p. 38).

## *Ecological concepts of information*

If we accept Gibson's (1966) claim that perception is based upon information in ambient arrays, and if we accept his assertion that perception is achieved by distinct perceptual systems, then we can ask about relations between types of information available to the distinct perceptual systems. Within the Ecological Approach to Perception and Action, this issue has been addressed in one of two ways. First, it often is claimed that specification of some aspect of reality exists within a given, single form of ambient energy, such that sensitivity to this particular parameter of a given ambient array would be sufficient accurately to perceive that aspect of reality. Examples include global optical flow (Gibson 1966), which is created by self-motion relative to an illuminated environment, and patterns in the haptic array produced by the inertia tensor (Solomon & Turvey, 1988), which is a property of handheld objects. We refer to this view as *modal specification* (Stoffregen & Bardy, 2001, Section 3.2). The principle alternative to modal specification is what we refer to as *amodal specification* (Stoffregen & Bardy, section 3.3.2), in which it is asserted that a given fact is equally, or redundantly specified in multiple forms of ambient energy. In the domain of speech perception, Fowler (2004) and Rosenblum, Dorsi, & Dias (2016) have articulated a position that is overtly amodal, arguing that speech gives rise to patterns in the optic and acoustic arrays that carry the same (specifying) information. In the context of development, a broader but otherwise nearly identical argument was presented by Eleanor Gibson (1983), who stressed the distinction between amodal information that is redundant across forms of ambient energy, and "modality-specific" information that is exclusive to particular forms of ambient energy.

The modal and amodal specification views echo Aristotle's (1931) distinction between the special objects and the common sensibles. Modal specification, in which a fact is uniquely related to a parameter of a single form of ambient energy (such as color, or pitch), corresponds to Aristotle's concept of special objects of perception, that is, percepts that are available exclusively to one sense. Amodal specification, in which a fact is redundantly related to parameters that exist in different forms of ambient energy (such as speech, movement, or number), corresponds to Aristotle's concept of the common sensibles.

## Conflict between perceptual systems?

Gibson (1966) asserted that some facts are specified exclusively within a single form of ambient energy (Aristotle's special objects of perception), and that other facts are specified redundantly in multiple arrays (Aristotle's common sensibles). Both Gibson and Aristotle gave examples of common sensibles, but neither offered a principled basis by which we might determine exactly those things that are (or can be) common sensibles. In Gibson's theory, the concept of common sensibles implies that patterns in different forms of ambient energy are redundant or equivalent. This idea of cross-modal redundancy has been adopted widely within the ecological approach (e.g., Fowler & Dekle, 1991; Gibson & Walker, 1984; Rosenblum & Saldaña, 1996). Stoffregen and Bardy (2001) argued that redundancy across perceptual systems implies the comparison, within the perceiver, of information available to individual perceptual systems, that is, cross-modal comparison. Cross-modal comparison appears to be required, for it is only through such comparison that we can differentiate things that are redundantly specified from things that are not. What happens, then, when the thing being perceived does not redundantly structure different forms of ambient energy? In such cases, the information available to different senses would be different, rather than redundant. That is, patterns available to one perceptual system (e.g., such as optic flow) might differ from patterns available to another perceptual system (e.g., acoustic silence).

Such differences are not compatible with the concept of lawful, 1:1 specification because any comparison of patterns in different forms of ambient energy must yield discrepancies. This argument was made at length by Stoffregen and Bardy (2001, section 3.3.3). Ordinary behavior always yields differences between patterns available in different single-energy arrays (e.g., Oman, 1982; Stoffregen & Riccio, 1991). As two examples, consider walking, and speech.

Walking typically structures light (as the head moves relative to the illuminated environment), sound (as the feet strike the ground surface) and gravitoinertial force (as the feet strike the ground and the head is positively and negatively accelerated). Speech commonly structures sound (emanating from the speaker's vocal tract), and light (reflecting off the speaker's face).

Walking and speech influence the structure of multiple forms of ambient energy. How can we describe and understand this fact? One way is to adopt a reductionist approach, to reduce the potential sensory stimulation resulting from a perceivable unitary event into constituent parts. The most common form of this reductionist approach is to examine the ways in which unitary events influence the structure of individual forms of ambient energy. This reductive analysis (which is consistent with the assumption of separate senses) leads to descriptions of potential sensory stimulation in terms of discrete patterns that exist in individual forms of energy (Figure 3).

As shown in Figure 3, many ordinary behaviors lead to patterns in single energy arrays that are different, such that comparison between single energy arrays must yield discrepancy, rather than redundancy. In the case of walking, footfalls yield a series of punctate compressions of receptors in the soles of the feet. But footfalls yield a different pattern in stimulation of the vestibular system: Due to the physical dynamics of energy propagation through the body (in this case, from the feet, upwards), changes in gravitoinertial force at the vestibule are not punctate, but more oscillatory in nature. Finally, the optical consequences of walking vary as a function of the 3-dimensional layout of the environment, which has no influence on stimulation of pressure receptors in the feet, or of the vestibule. Similarly, in speech, the acoustic waveform that exits the mouth is influenced by but differs from the visible activity of the face (i.e., jaw, lips, and facial

muscles).³ Speech generally does not alter patterns of gravitoinertial force at the point of observation (i.e., a listener), such that patterns in gravitoinertial force will be entirely independent of (and different from) patterns in optics and acoustics. In each case (walking, and talking), comparison of patterns that exist in individual forms of ambient energy must yield discrepancies: As shown in Figure 3, the patterns in the individual forms of ambient energy are simply different, so that no pattern in any individual form of energy is uniquely related to reality. For this reason, lawful, 1:1 specification cannot exist in single energy arrays (see Stoffregen & Bardy, 2001, for a complete presentation of this argument). Thus, to preserve the concept of lawful, 1:1 specification (and, therefore, to preserve the possibility that perception might be direct) we must move away from the Aristotelian concepts of modal and amodal sensation, and from Ecological concepts of modal and amodal specification (Fowler, 2004; E. J. Gibson, 1983; J. J. Gibson, 1966; Rosenblum et al., 2016).

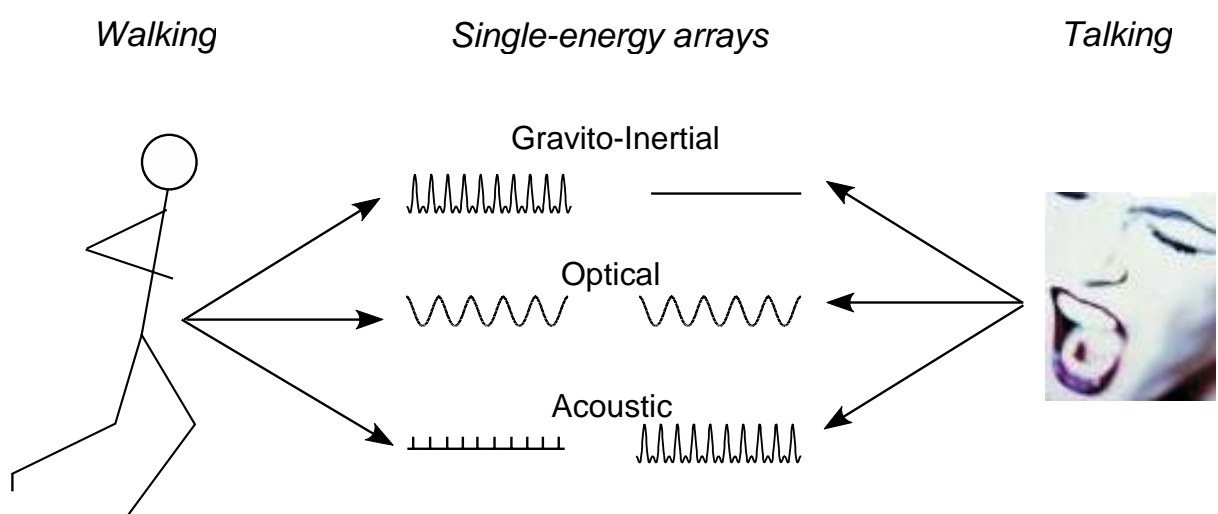

**Figure 3.** Unitary events give rise to patterns in single-energy arrays that are different. There is difference between patterns created in forces, optics, and acoustics by walking. Similarly, there is difference between patterns created in forces, optics, and acoustics, by talking.

Perception can be understood as a type of measurement. Like any form of measurement, it must be conducted relative to some referent. To perceive, for example, that something is three feet long entails the idea that perception of length is scaled in terms of units of feet. Similarly, to perceive that an object is within reach entails the idea that perception is scaled in units of arm length or, more broadly, of reachable distance. What are the referents for perception? In classical, inferential theories of perception, referents for perception are internal to the perceiver; often known as *sensory reference frames* (e.g., Soechting & Flanders 1992; Stoffregen & Bardy, 2001, Section 3.1.2). By contrast, if perception is direct (as asserted in the Ecological Approach to Perception and Action), then the referents for perception can be external to the perceiver, that is, they can be facts of the animal-environment system. In the present article, we accept as given the idea of direct perception, based upon lawful, 1:1 specification. Accordingly, we take as given the idea that referents for perception are "outside the head", located in the animal-environment system (cf., Mace, 1977).

---

³ The fact that comprehension is improved when the optical and acoustical consequences of speech are simultaneously available is tangential to this point. Related issues, such as the McGurk effect, were discussed by Stoffregen and Bardy (2001, Sections 3, 6.2.6).

Analyses of perception typically focus on individual facts about the animal-environment system. How far can I reach with this rod (e.g., Solomon & Turvey, 1988); can I catch this fly ball (e.g., Oudejans et al., 1992); can I (or you) sit on this chair (Mark, 1987; Stoffregen, Gorday, Sheng, & Flynn, 1999), and so on. Stoffregen & Bardy (2001; sections 4.4, 4.5) introduced the concept of *multiple simultaneous referents*. They argued that in most natural situations, different aspects of behavior are simultaneously controlled relative to qualitatively different referents. That is, they argued that perception is rarely of "a single thing" (cf., Wagman, Caputo, & Stoffregen, 2016). If perception is for action, then perception should provide information about aspects of the animal-environment system that are relevant to action. To reach (e.g., with a rod) certainly depends upon the length of the rod, but it also depends upon the perceiver/actor's ability to achieve and maintain an upright posture while holding and wielding the rod. This is an integrated skill that is learned in infancy, and has been studied and documented extensively (e.g., Thelen & Spencer, 1998). Gibson (1966) was explicit in stating that all "higher" action (including stable perceptual contact with the surroundings) was fundamentally dependent upon the ability to perceive and control (that is, to stabilize) the body. Our own actions generate forces that alter the constraints on control of the body. Stoffregen and Riccio (1988) detailed extensively the fact that animate movement changes, in a highly dynamic manner, the forces acting on the body, and against which the body must be stabilized. To wield a tool is to displace not only the mass of the tool but also the mass of the body (for example, the arms), in ways that alter the magnitude and direction of the Direction of Balance, which is contraparallel to the vector sum of gravity and inertial forces (Riccio & Stoffregen, 1990). Examples are given in Figures 4-6.

Due to these physics, specification does not and cannot exit in single energy arrays; they are intractably ambiguous with respect to the animal-environment interaction, that is, to the functional contents of perception. One way to understand our claims is to review some of Gibson's (1966) arguments about specification. In the next section, we review and critique Gibson's concept of the *point of observation*. In a subsequent section, we propose a new concept of the point of observation.

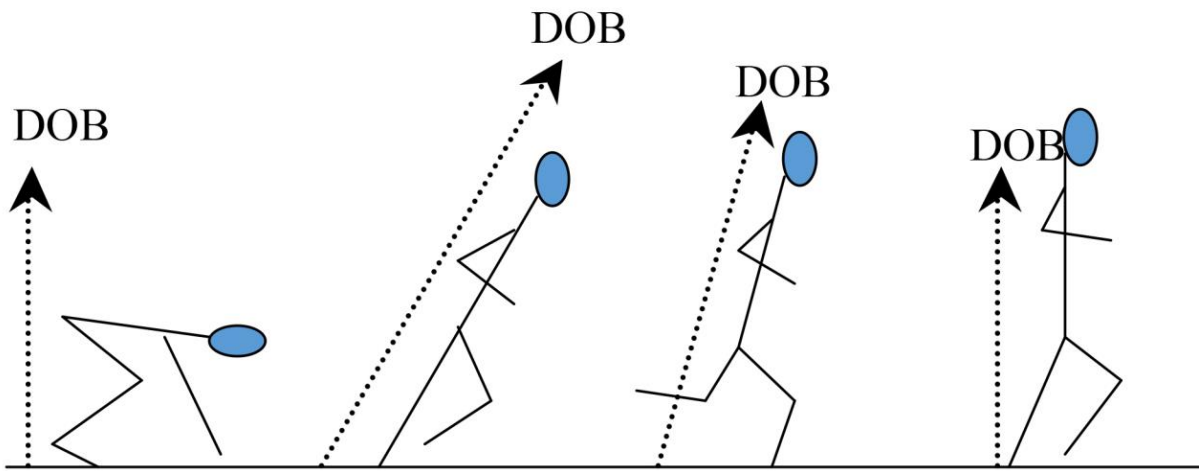

**Figure 4.** Multiple, simultaneous referents: Body orientation must be controlled relative to the Direction of Balance (DOB, the direction contraparallel to the vector sum of gravity and inertial forces), which is dynamic, while body translation must be controlled relative to the ground surface, which is static.

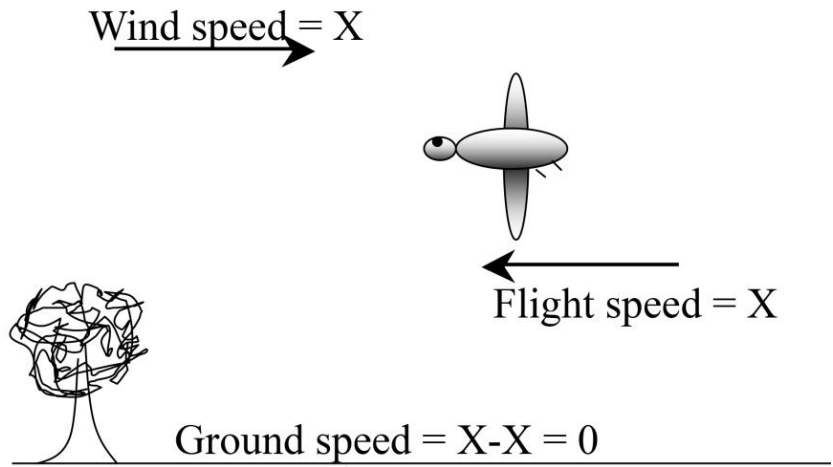

**Figure 5.** Multiple simultaneous referents for non-humans. The bird must control aerodynamics relative to the air mass, but often does this so as to maintain some trajectory relative to the ground. The air moves relative to the ground, yielding multiple simultaneous referents. In a high wind, a bird can be flapping vigorously but remain stationary relative to the ground. A comparable case obtains for human pilots.

## The disembodied point of observation

Gibson's (1950, 1966) offered the *point of observation* as an aid to analyzing patterns in ambient arrays and, more broadly, as a way to advance his arguments about lawful, 1:1

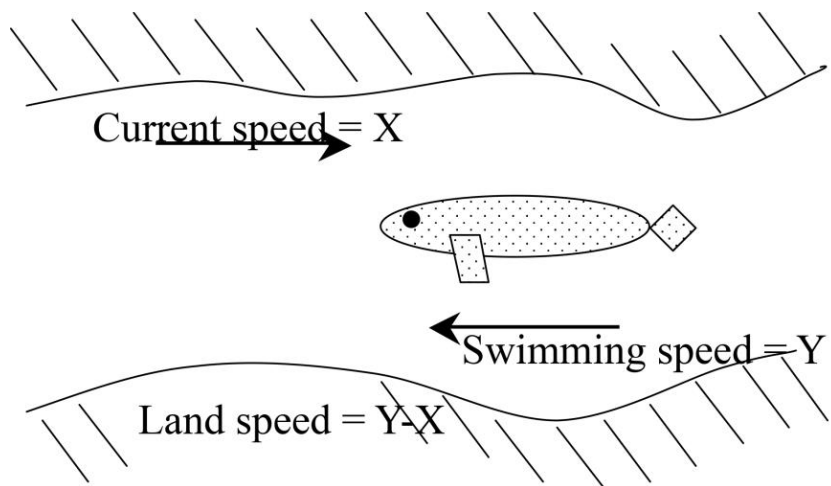

**Figure 6.** Multiple simultaneous referents for non-humans. The fish controls hydrodynamics relative to the water, which may be in motion relative to the land (or things in the air, e.g., insects). To maintain perceptual contact with things that are stationary relative to the land (e.g., to swim toward a rock), the fish must control its motion relative to the current (the water in motion), and simultaneously relative to the illuminated environment. The same case obtains for human swimmers.

13/34

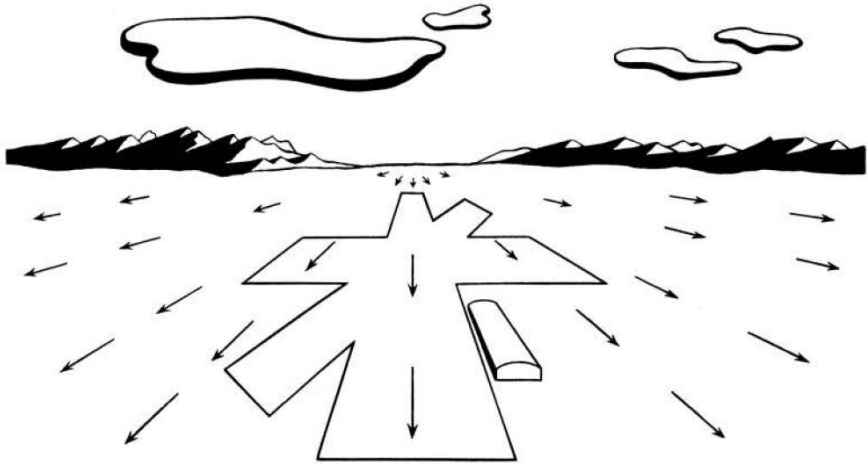

FIGURE 53. Motion Perspective in the Visual Field Ahead

**Figure 7.** Optic flow at a disembodied point of observation. Reproduced from Gibson (1950; Figure 53).

specification. Originally, Gibson's concept of the point of observation was developed in the context of vision. The classic illustrations of optic flow, reproduced here (Figure 7 and 8), were developed from powered flight, in which motion of the point of observation was mechanical, such that the embodiment of the observer was relatively unimportant: The (ground-relative) kinematics of optic flow came from the aircraft, not from the observer's self-generated locomotion. Gibson's purpose was to illustrate the abstract concept of optic flow, and this was easily done by ignoring the kinetics, the physical properties, the embodiment of the observer. The idea that the kinematics of motion can be separated from the kinetics of motion (in effect, the idea that kinematics and kinetics can be understood as distinct, qualitatively different components of dynamics) is arbitrary; an analytic convenience. The fact that kinematics can be separated from kinetics for analytic purposes (e.g., by behavioral scientists, or by physicists) does not imply that kinematics and kinetics actually are separate, and separable. The separation of kinetics from kinematics is an instance of reductionism, and it can succeed only if in this separation nothing is lost. That is, the analytic separation of dynamics into kinematics and kinetics entails the assumption that dynamics is nothing more than the additive sum of kinematics and kinetics.

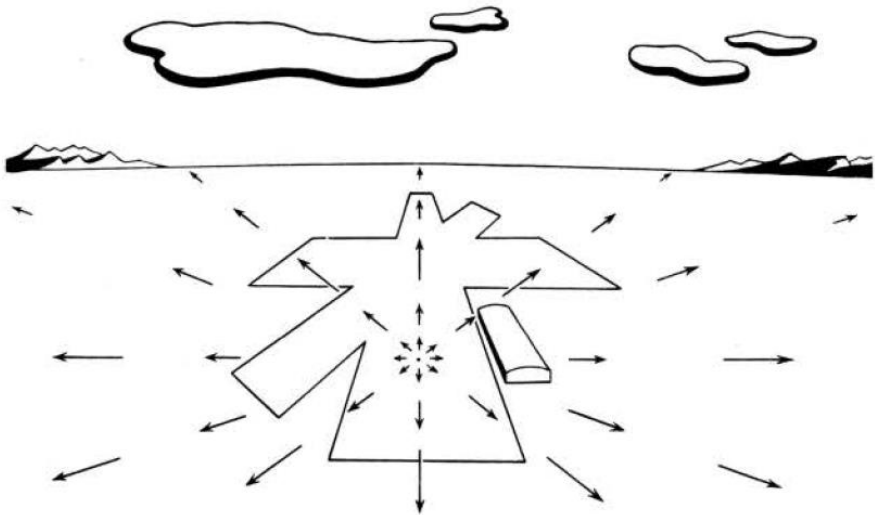

FIGURE 58. Gradients of Deformation during a Landing Glide

**Figure 8.** Optic flow at a disembodied point of observation. Reproduced from Gibson (1950; Figure 58).

The disembodied illustrations of Figure 7 and 8 are reflected in the equations that provided the original mathematical formalization of optic flow (Gibson, Olum, & Rosenblatt, 1955). In these equations, all of the terms were kinematic; positions, angles, velocities. There were no kinetic terms. Points of observation were geometric points, not physical locations. They were empty points in space, not points occupied by physical observers. In 1955, this approach made sense because the authors were making a claim about "pure" optics, in which kinetics can play no part. Our discussion makes clear that any analysis of "pure" optics necessarily entails a disembodied point of observation; a geometric point, rather than a physical thing in a physical world.[4]

Gibson et al. (1955, p. 372) described motion parallax in terms of the orbital motion of planets. The use of a disembodied point of observation was derived from astronomy, but also served to enable formalization of equations for "pure" optic flow (e.g., Gibson et al., 1955). We argue that "pure" optic flow is disembodied optic flow. In general, single-energy arrays are ambient in the disembodied sense described in 1955, and in 1966.

## *The embodied point of observation*

We argued, above, that the separation of kinematics from kinetics in the original formulation of the optic array was arbitrary. Here, we argue that analysis of patterns in ambient energy must be based on what we call the *embodied point of observation* (Stoffregen, 2014). That is, we argue that any successful analysis of lawful, 1:1 specification must include the full dynamics (that is, kinetics and kinematics), and not merely the kinematics of the point of observation.

The embodied point of observation is not a point, in the geometrical sense. It is a region of space, having size, dimensions, and dynamics corresponding to the size, dimensions, and dynamics of the animal. The position and motion of the embodied point of observation are influenced by properties of the animal that inhabits that point of observation. As one example, the quantitative kinematics of standing body sway vary depending upon a person's age, height, weight, sex, and foot length (Chiari, Rocchi, & Cappello, 2002; Kim et al., 2010), their foot position (Stoffregen, Yoshida, Villard, Scibora, & Bardy, 2010), athletic or dance skill (Schmit, Regis, & Riley, 2005), state of health (Schmit et al., 2006), position in the menstrual cycle (Darlington, Ross, King, & Smith, 2001), fatigue (Vuillerme, Baptiste, & Rougier, 2007), and many other factors. That is, properties of the living, physical individual lead to differences in movement. Whole body movement, including body sway, alters the stimulation of the visual, auditory, haptic, and vestibular systems, and has been implicated as a source of information about affordances (Mark et al., 1990; Stoffregen, Yang, & Bardy, 2005; Yu, Bardy, & Stoffregen, 2011). This concept is not limited to body sway, but applies equally to movement of all kinds, at all scales. The kinematics of gait are influenced by physical properties of the body, including weight (Lai, Leung, Li, & Zhang, 2008), clinical conditions (Morris et al., 2001), and sex (Kerrigan, Todd, & Croce, 1998), and by non-physical properties of the animal, such as skill (Teplá, Procházková, Svoboda, & Janura, 2014), or recent history, such as sea travel (Stevens & Parsons, 2002). Manual activity differs between individuals (Tretriluxana, Gordon, & Winstein, 2008), which implies that the information generated by hefting and wielding will also differ. The patterns in ambient energy created by whole body movement are so distinctive that they are sufficient for us to recognize our friends (Loula, Prasad, Harber, & Shiffrar, 2005), and to yield individual "signatures" of movement (Slowinski et al., 2016). The learning of new patterns of coordination in whole body movement differs between individuals (e.g., Faugloire, Bardy, & Stoffregen, 2006), which implies that the information generated during learning will be similarly unique, or embodied.

---

[4] Our comments apply only to the equations in Gibson et al. (1955). The authors acknowledged that their equations omitted necessary information about kinetics (p. 384). Subsequent efforts to identify information in optics have rarely been so frank about the fundamental incompleteness of the effort (e.g., Lee, 1980; cf. Bingham & Stassen, 1994).

Any situation in which available information is structured by activity of the perceiver is a situation that alters the structure of multiple forms of ambient energy, whether the scientist chooses to recognize and analyze that fact, or not. Therefore, any point of observation that is occupied will be a point that has position and motion relative to both kinematics and kinetics; that is, relative to multiple simultaneous referents. This is ecological physics, and it is logically prior to specification. The embodied point of observation can be inferred from the concept of the reciprocity of animal and environment (Lombardo, 1987), extending that concept to the level of quantitative properties of the animal, including quantitative properties of movement that generates information.

In terms of perceptual information, Gibson (1966, p. 22-23) was clear about this: "the kind of light it reflects, the kind of sound it makes, and the kind of chemical it diffuses will all specify the sort of animal it is—carnivore or herbivore, male or female, young or adult." This logic extends beyond the stated categories to the uniqueness of the individual, including the uniqueness of individual movement, that is, to the embodied point of observation.

We accept that many facts about the world are specified at multiple points of observation, a point made often by Gibson (1966). The fact that, for example, a child is asleep is specified in patterns that are available at many locations, and along many trajectories. Yet at the same time, the same patterns also specify the location and trajectory of those points of observation and, in our argument, also specify the occupant of any point of observation (e.g., a parent checking on their sleeping child), that is, the dynamics of the point of observation, rather than solely its kinematics. In essence, our argument about the embodied point of observation is simply the logical limit of the concept of animal-environment reciprocity (Lombardo, 1987), with the essential proviso that the point of observation is always occupied by a living thing with mass as well as position and that, therefore, the point of observation always is the nexus for flux in multiple forms of ambient energy.

An early empirical example of the embodied point of observation comes from Mark et al. (1990), who showed that active movement was sufficient for perception of an affordance. Standing participants were asked to judge their own maximum sitting height. Mark found that judgments were grossly inaccurate when participants were deprived of all bodily movement (e.g., by being required to stand with their back and head held against a wall during judgments), but that something as minimal as ordinary body sway provided movement sufficient for judgments to be accurate. The results of Mark et al. imply the embodied point of observation. The dynamics of optic flow generated by sway would be influenced by the dynamics (kinetics) of the body; hence, the physical body was implicit in the fact that sway was sufficient. Yet, the fact that the body influences the optical consequences of sway does not reciprocally imply that these optics are sufficient to specify body sway, or the body. This is so because the identical optics can be generated without a body, without an animal; for example, by playing back a video recording of the optical consequences of actual sway (this issue is addressed in detail in a later section). The fact that the same optics can be generated by two different events (a body in motion, or a video recording of a body in motion) means that the optics are not specificational. However, when a body sways, the optic flow generated by sway co-occurs with sway-created changes in other forms of ambient energy (Stoffregen & Bardy, 2001, Section 6.1; cf. Bingham & Stassen, 1994).

## The global array and the perceptual system

Given that organisms need information about the animal-environment interaction, where can they get it? Are there patterns of ambient energy that contain the information that is needed? If so, then evolution should lead to perception that is sensitive to just these patterns, to just this information.

### *The global array: Patterns in ambient energy*

Many forms of energy exist, and are ambient to organisms. These forms of energy exist simultaneously. As noted above, because living things have mass and occupy space, the point of observation is embodied, such that behavior alters the structure of multiple forms of ambient energy. Traditional analyses of information have been limited to parameters that exist in individual forms of ambient energy. Until recently, the sole exception was Gibson, 1966, Chapter IV (as described above; Figure 1), who proposed that information could exist as emergent, higher order patterns that extended across different forms of ambient energy. Stoffregen and Bardy (2001) argued that such emergent, higher order patterns are not merely possible, and do not exist in a merely occasional sense. Rather, they asserted that emergent, higher order patterns extending across multiple forms of ambient energy are, in principle, the *only* type of patterns that meet the criteria for lawful, 1:1 specification of the animal-environment system. The set of patterns extending across multiple forms of ambient energy was named by Stoffregen and Bardy the *global array*. The global array consists of patterns that extend across multiple forms of ambient energy, and which exist solely as emergent, higher order relations between multiple forms of ambient energy. Qualitative examples are given in Figure 9 (compare with Figure 3).

Single-energy arrays, such as the optic array and the acoustic array, are compatible with the traditional concept of the disembodied point of observation, and are not compatible with our concept of the embodied point of observation. The global array is compatible with the embodied point of observation. The global array is not ambient to massless, geometric points; rather, it intrinsically, inescapably is ambient to and provides information about the full dynamics of the animal-environment interaction. The global array pulses with the life of the organism; it embodies animal-environment reciprocity. In this sense, we emphasize that animal-environment reciprocity is between the environment and the animal (e.g., Lombardo, 1987), not between the environment and purely kinematic entities, such as geometric station points, or trajectories.

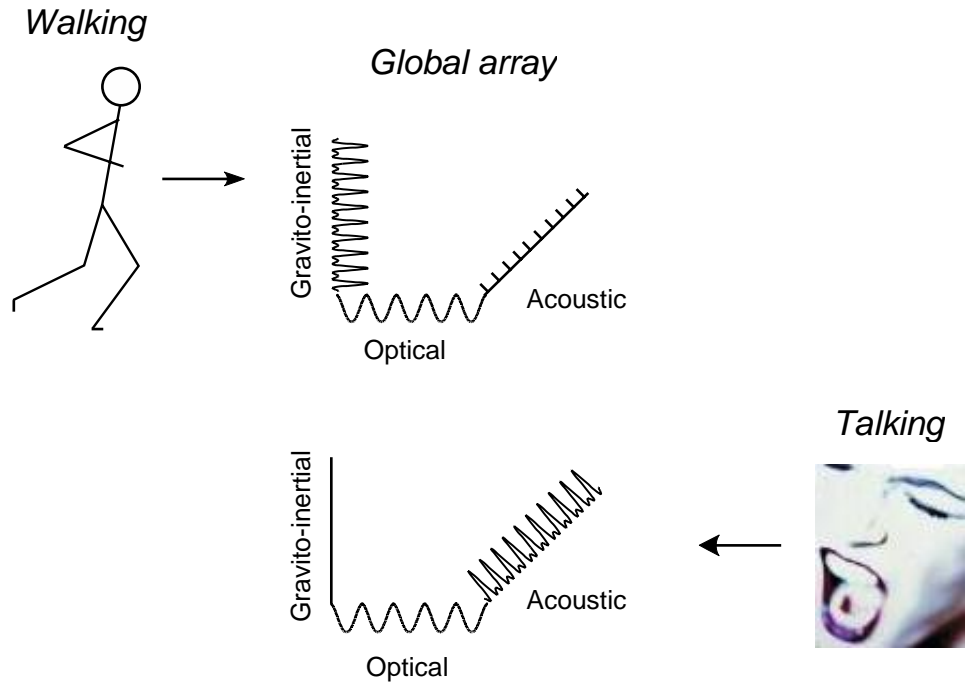

**Figure 9.** Parameters in the global array exist as emergent, higher order patterns that extend across different forms of ambient energy. Compare with Figure 3.

## The global array: Specification

Gibson argued that information (specificational parameters in ambient energy) can exist as (higher order) invariant relations between other (lower order) invariants: "a unique combination of invariants, a *compound* invariant, is just another invariant. It is a unit, and the components do not *have* to be combined or associated," (Gibson, 1979/1986, p. 141, emphasis in the original). Parameters of the global array are nothing more and nothing less than compound invariants. They differ qualitatively from the lower order, constituent invariants that they comprise. This qualitative difference means that compound invariants do not exist in, and cannot be derived from the lower order component invariants. Parameters of the global array emerge from relations between patterns in single-energy arrays; they are irreducible.

Because multiple referents always exist, and because behavior relative to these referents always influences the structure of multiple forms of ambient energy, it is possible that specifying patterns (emergent properties) extend across multiple forms of energy, such that specification is neither modal nor amodal, but exists at a higher order. Patterns in the global array are always available. If behavior always entails the simultaneous control of the animal-environment interaction relative to multiple, simultaneous referents (as expressly claimed by Gibson, 1966, Chapter IV), then information about the position and motion of the animal relative to multiple, simultaneous referents is always needed.

The existence of specification in the global array is not in addition to specification in single-energy arrays. Rather, the two possibilities are mutually exclusive. Stoffregen and Bardy (2001) argued that lawful, 1:1 specification exists solely in the global array, and that all patterns in single-energy arrays are intractably ambiguous with respect to the dynamics of the animal-environment interaction.

If it is accepted that the global array exists, and if it is accepted that the global array contains information that does not exist in single-energy arrays, then we must confront the question of how information from the global array might be picked up. We propose that the senses form a single, unitary, irreducible perceptual system whose sole function is to detect parameters of the global array. Many students of the Ecological Approach to Perception and Action are used to the assumption that perception happens in overlapping but discrete systems, and so our proposal that perception happens solely through a single, unitary system may appear strange. We do not propose that the various perceptual systems "work together" to obtain parameters of the global array. Such a proposal would imply that there *are* separate systems that might or might not work together (e.g., relative to the common sensibles) but might also work separately (e.g., relative to the special sensibles). We reject this view. Our argument is that there is only one perceptual system, and that it always and only functions as a single system.[5]

Gibson (1966) understood that information consisted of emergent, higher order patterns, that is, relations between lower order elements in the structure of ambient arrays. In an earlier section, we raised this point in the context of optic flow. But this point extends more broadly, in ways central to our argument. In terms of structure in optics, the two eyes form a single system. Working together, the eyes are sensitive to parameters of the optic array that are related to spatial layout (i.e., depth), and which exist exclusively as higher order relations between simultaneous pairs of optic array samples. That is, certain aspects of spatial layout in the illuminated environment can be detected only at the emergent level of the binocular system. Similarly, in terms of acoustic structure, the two ears form a single, binaural system (Gibson, 1966, Chapter V), which is sensitive to ambient patterns (e.g., patterns related to egocentric location of sound sources) that can be detected only at the level of the binaural system. In both cases, the information does not exist in patterns available to one sensor (one eye, or one ear), but exists (and can be sampled) only at the level of irreducibly singular function of anatomically distinct sense organs. If these examples are accepted as real, then the conceptual leap to the detection of parameters of the global array is very small.

"The detecting of stimulus information without any awareness of what sense organ has been excited, or of the quality of the receptor, can be described as 'sensationless perception'" (Gibson, 1966, p. 2). We assert that perception through the global array rarely, if ever, leads to conscious sensation in the classical sense to which Gibson objected. Perceptual systems, Gibson asserted, are "ways of paying attention to whatever is constant in changing stimulation," (p. 4). Similarly, "What might be a physiological or functional equivalent of the external information, if it cannot be anatomical? How could invariants get into the nervous system? The same incoming nerve fiber[s] makes a different contribution to the pickup of information from one moment to the next. The pattern of the excited receptors is of no account; what counts is the external pattern that is temporarily occupied by excited receptors…" (p. 4). In these quotations, Gibson did not refer to distinct perceptual systems. We assert that in these quotations, Gibson was (unintentionally) presaging our concept of the perceptual system.

---

[5] As noted by Stoffregen & Bardy (2001, Sections 6.2.3. and R11), sensory loss (e.g., blindness) affects the *pickup* of information from the global array, but has no effect on the *existence* of information in the global array.

19/34

"The function of the brain is not even to *organize* the sensory input or to process the data, in modern terminology. The perceptual systems, including the nerve centers at various levels up to the brain, are ways of seeking and extracting information about the environment from the flowing array of ambient energy", (p. 5). By changing the plural to singular in this quotation, we have a single perceptual system working to extract information from the global array. "Since the senses are being considered as perceptual systems, the question is not how the receptors work, or how the nerve cells work, or where the impulses go, but how the systems work as a whole. We are interested in the useful senses, the organs by which an organism can take account of its environment and cope with objective facts", (p. 6). Again, by changing "perceptual systems" to "a perceptual system", this quotation captures our view.

Classically, it has been assumed that patterns that exist in individual forms of ambient energy are separately detected by distinct senses (or perceptual systems), and that relations between these patterns are determined or identified within the nervous system, through some process of integration. Stoffregen and Bardy (2001) argued that, within the Ecological Approach to Perception and Action, this view leads to discrepancies (typically interpreted as intersensory conflict) that are inconsistent with the concept of lawful, 1:1 specification. The existence of emergent, higher order patterns in the global array nullifies this issue, "for the neural inputs to [the] perceptual system are *already* organized and therefore do not have to have an organization imposed upon them … the available stimulation surrounding an organism has structure … and … this structure depends on sources in the outer environment. If the invariants in this structure can be registered by [the] perceptual system, the constants of neural input will correspond to the constants of stimulus energy… The brain is relieved of the necessity of constructing such information by *any* process" (Gibson, 1966, p. 267, emphasis in the original).

## *Simulation and differentiation*

Experimenters and technologists often attempt to simulate various types of perceptual information. There is no rigorous, widely accepted definition of *simulation*.[6] In the context of perceptual theory, we can offer a qualitative definition: A simulation is an attempt to present patterns in ambient arrays corresponding to some particular physical event but in the absence of the physical event. Under this definition, the cinema constitutes a form of simulation. In the cinema, physical events are acted out in front of a camera. The camera records the optical patterns arising from the physical events. Later, these optics can be displayed in the absence of the original, physical event; usually, this is done by projecting the recorded optics onto a screen. People who view these recordings recognize the physical events that took place in front of the camera (e.g., Anderson, 1996). With contemporary technology, physical events are no longer required. Through digital animation, we can create static and kinematic optic patterns in the absence of any underlying physical events. Using simulation technologies, we can create optics that correspond to physical events that have never happened (e.g., George Washington shaking hands with King George III), and even for events that cannot happen (e.g., people growing old "backwards", as occurred in *The Curious Case of Benjamin Button*). These simulations look real, even when the viewer knows that they do not correspond to actual, physical events. Effects of this kind have led to considerable confusion about what is perceived when we view the cinema, or any visual simulation, including contemporary computer graphics (Anderson, 1996; Stoffregen, 1997). In these simulations, the optics exist free of the physical universe. They are disembodied, in the same sense as the disembodied point of observation introduced by Gibson et al. (1955).

---

[6] Issues of simulation, fidelity, presence, and immersion extend back to classical antiquity, perhaps most famously in the myth of Amphitryon (e.g., Dupuy, 2008). The logical problems inherent with simulation in the context of experimental stimuli and virtual reality are not new.

Is cinema perceived as reality? Does the sense of "immersion" that many people feel in virtual environments mean that we cannot distinguish virtual reality from the physical world? We argue that the answer to these questions is categorical: No. Viewers who are impressed by the realism of cinema, or of virtual reality nevertheless readily differentiate the virtual from the physical (for an extended discussion of perceptual experience in the cinema, see Stoffregen, 1997). If we accept that observers accurately differentiate simulation from reality, we can ask about the perceptual basis for such differentiation. Do virtual and physical worlds give rise to different patterns in ambient arrays and, if so, what is the nature of the differences? Or, to put the question differently, is simulation specified, and if so, how? These questions are of interest for general, theoretical reasons, but also because simulations are commonly used as stimuli in experimental research on perception (e.g., Fajen, 2005).

Observer movement relative to a flat screen (e.g., a projection surface for cinema or virtual reality) yields patterns in the global array that are different from patterns generated by observer movement relative to a physical, 3-dimensional layout (Stoffregen, 1997). This point was made originally by Gibson (1966, 1979), and remains valid. The same point applies, in slightly subtler form, to the consequences of observer movement relative to "3-D" simulations. In essence, this is the issue of simulation fidelity: Can a simulation ever be indistinguishable from the corresponding physical reality? We have argued that for active, unconstrained observers, simulations can always be distinguished from the corresponding physical reality (Stoffregen, Bardy, Smart, & Pagulayan, 2003). That is, we have argued in formal terms there is no simulator fidelity: The fact of simulation is specified. This argument applies not only to optic simulations, but to simulations in other single-energy arrays (e.g., acoustics, as in recorded, or synthesized sound, including speech). It applies equally to "multimodal" simulations, as illustrated in Figure 10 (compare with Figure 9). The fact that experimental participants can perceive simulations (i.e., can recognize the thing or activity that is being simulated), and can control simulations does not imply that the information detected and used under experimental conditions is the same as in the outside world (e.g., Fajen, 2005). We are not suggesting that simulations can never be used in experimental research on perception and action. Rather, we claim that the fact of simulation is specified in global array patterns that are available to experimental participants and that, depending upon the hypotheses being tested, this fact may have consequences for the interpretation of experimental data.

## Gibson's legacy

In an earlier section, we raised broader issues concerning the precise meaning of Gibson's (1966) arguments. We have argued that Gibson did not explicitly reject the Aristotelian distinction between common sensibles and special objects of perception. It is not easy to make an argument based on what a scholar did *not* say. However, we can examine claims that have been made by later scholars who were influenced by Gibson. In certain important ways the impact of Gibson's book has been broadly consistent. As noted by Stoffregen and Bardy (2001, Section 3.3), many acolytes of Gibson have adopted some version of the idea that certain facts are available only to a single perceptual system while others are "equally", or "redundantly" available to multiple perceptual systems. Examples cited by Stoffregen and Bardy included Bahrick (1988), Fitzpatrick, Carello, Schmidt, and Corey (1994), Lee (1990), and Rosenblum and Saldaña (1996).

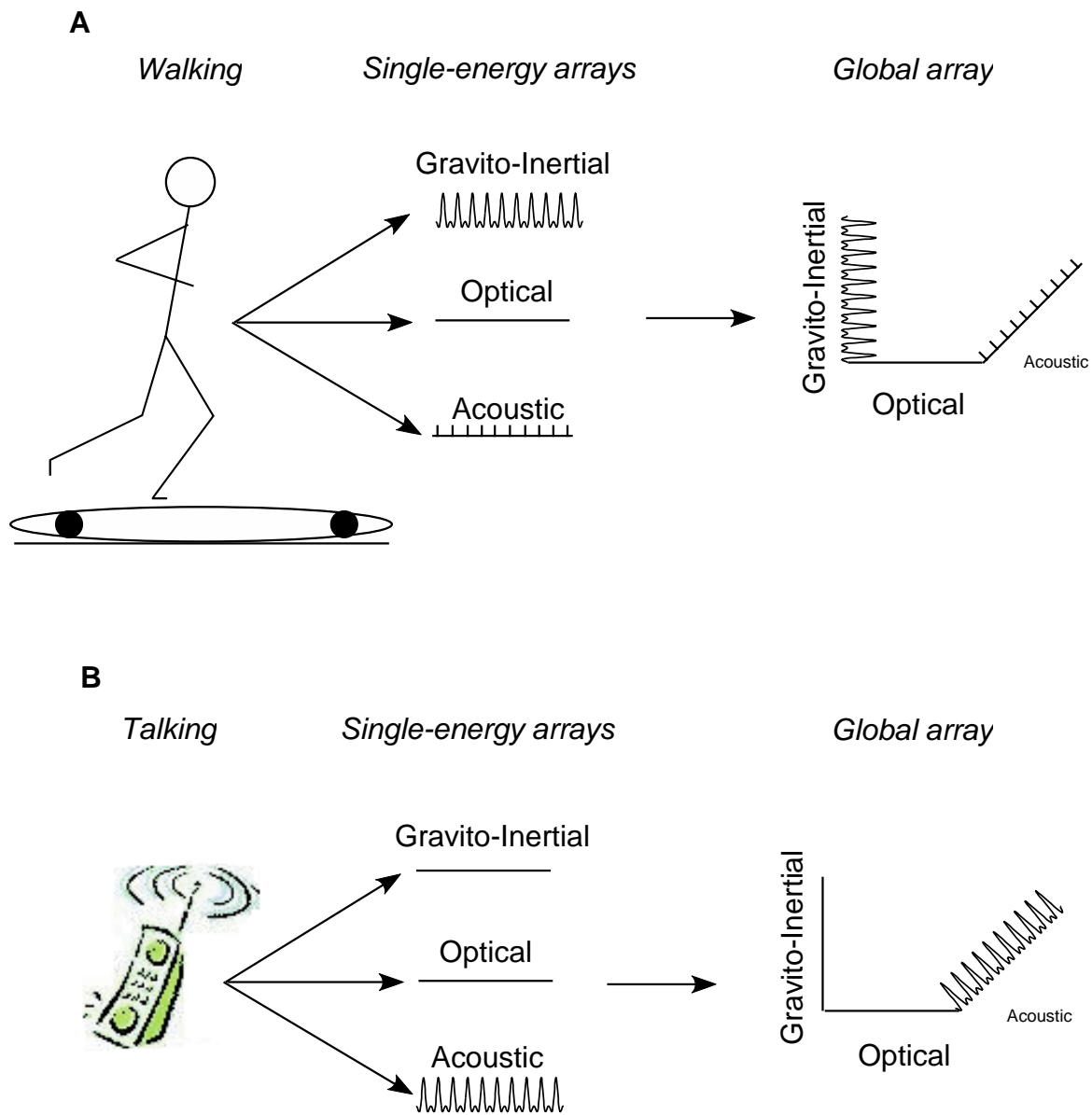

**Figure 10.** Simulations yield patterns in the global array that are different from patterns generated by the corresponding physical animal-environment interactions. The telephone counts as a simulation because the sound is reproduced through the telephone, rather than coming directly from the human speaker. Compare with Figures 3 and 8.

Within the Ecological community, the common sensibles have been appealed to—explicitly, or implicitly—in many analyses of situations that involve simultaneous stimulation of multiple perceptual systems. One current example is the literature on speech perception. Fowler (2004; cf. Rosenblum et al., 2016) proposed an essentially Aristotelian view of speech perception. She pointed out, in great detail, the fact that speech alters the structure of multiple forms of ambient energy, and she reviewed empirical evidence demonstrating that the perception of speech not only can be but routinely is influenced by non-acoustic stimulation. She reviewed a wide variety of theories of speech perception, each of which made some attempt to account for the fact that speech perception is influenced by non-acoustic stimulation. The legacy of Gibson (1966) can be identified in a single sentence: "structure in informational media (e.g., light for vision, acoustic signals for hearing) is caused by properties of environmental events" (Fowler, 2004, p. 198). In this sentence (and throughout Fowler's argument) we see an implicit assumption that information exists in different forms of ambient energy, and that perception acts to detect information in individual forms of ambient energy. Indeed, the concept of amodality necessarily entails the concept of modality, that is, of multiple, distinct modalities. Only under that assumption (derived from Gibson and, ultimately, from Aristotle) does the concept of amodality have coherence. Stoffregen and Bardy (2001) made this same argument (using different examples) in their Section 3.3.2.

Fowler (2004) argued for the existence of a "common currency" in speech perception, and that this common currency was information (in the Ecological sense) that was "modality neutral". Rosenblum et al. (2016), elaborating Fowler's position, argued that "information can be composed of dynamic patterns of energy whose abstract nature allows it to be instantiated in multiple energy arrays while keeping its same form" (p. 264), such that the "common currency" is "shared between modalities" (p. 283). As with Gibson (1966) these authors do not explicitly address the relation between their arguments and the Aristotelian position. However, as written, their arguments appear to be consistent with the Aristotelian concept of common sensibles[7], that is, while multiple senses may have access to the same "distal events", they remain separate senses, and no appeal is made to the possible existence of information in emergent patterns that extend across different forms of ambient energy. Rosenblum et al. particularly stressed the idea that information available to different perceptual systems is "modality neutral" because it is (they claimed) redundant across perceptual systems.

In effect, Fowler (2004) and Rosenblum et al. (2016) argue that the lines that comprise a triangle, while they may differ in color or thickness, all closely resemble each other as lines (e.g., in length). They do not consider the possibility that relations between the lines may yield emergent, higher-order patterns (e.g., triangles) that differ qualitatively from any properties of the lines, as such. One problem with this approach is that it depends almost entirely on the assumption of redundancy (see Rosenblum et al., p. 263-266, for a list of alleged redundancies across optic and acoustic consequences of speech). The existence of distinct perceptual systems, and the separate function of separate perceptual systems, is not questioned.

The analysis of Fowler (2004) and Rosenblum et al. (2016) comprises widely divergent examples, but omits some that are relevant to our claim that information exists solely in the global array. People often converse without being within one another's field of view. As a common example, consider a conversation between one person in the dining room, setting the table, and another person in the kitchen, preparing a meal. The gravitoinertial patterns available to each person are peculiar to their activities (setting, cooking, listening), and the optical patterns are unique, but in different ways (the visible interior of the separate rooms, neither of which includes structure relating to the other person). Patterns in the acoustic array, by contrast, are related to a voice coming from another room. Taken at the level of single-energy arrays, the patterns in force, optics, and acoustics are different: Any comparison between them must yield discrepancy, or intersensory conflict (Stoffregen & Bardy, 2001, Section 3.3.3). However, the higher-order relation between patterns extending across these forms of energy is specific to what actually is happening—conversation with a person who is out of sight in another room. Consider also talking with a person while standing back to back (e.g., Shockley, Santana, & Fowler, 2003). The conversants are in the "same place", but relations between sound, light, and force specify what actually is happening—talking with another person who is occluded by one's own body (which, of course, cannot occur with a geometrical point of observation). Even when standing side by side, dyads often converse while looking at something other than each other—any object of joint attention (e.g., Richardson, Dale, & Kirkham, 2007). In each of these examples, any comparison between patterns available in single-energy arrays must yield intersensory conflict and, therefore, are at odds with the concept of 1:1 specification (as well as with the idea of "common currency", or amodality). These examples illustrate the point that the physical circumstances of the embodied point of observation structure multiple forms of ambient energy. In these cases, the concept of "common currency" (Fowler, 2004; Rosenblum et al., 2016) across perceptual systems fails. When two people talk about a baseball game that they are watching, the optical consequences of speech are not sampled, that is, they are not part of the global array that is obtained, in Gibson's (1966) sense (see the following section on exploratory activity). The same is true for conversation with a person who is out of sight (e.g., in another room), and for conversation via technologies that exclude the optical consequences of speech (e.g., telephones; Figure 10B). In essence, our analysis of the global array raises new questions for speech perception. Rather than asking how the optic and acoustic consequences of speech may be equivalent, our analysis of the global array raises questions about how it is possible for us to differentiate situations such as conversing face-to-face, conversing side-by-side (or back to back; Shockley et al, 2003), conversing with a person in another room, and conversing via telephone. What is the information that enables us to differentiate these common situations?

---

[7] Rosenblum et al. (2016, p. 268) contrasted "modality neutral" consequences of speech with "sensory specific dimensions" of stimulation. This contrast closely resembles the Aristotelian distinction between common sensibles and special objects of perception and, accordingly, entails the assumption of separate senses.

### The global array in 1966

In Chapter IV, Gibson (1966) presented (in qualitative terms) a parameter of the global array (Figure 1). Gibson (p. 62) noted, "[a]ll animals orient to [the gravitoinertial force environment] but terrestrial animals also orient to the surface of support." Orientation relative to the gravitoinertial force environment structures stimulation of the vestibular system, while orientation relative to the surface of support structures stimulation of the haptic system. When the ground is horizontal, the directionality of stimulation of the vestibular and haptic systems will be parallel. When the ground is not horizontal, the directionality of stimulation of the vestibular and haptic systems will be different (i.e., not parallel). The relation between these two directions is an emergent property that does not exist in stimulation of the vestibular system or in stimulation of the haptic system. That emergent, relational property "is information" (Gibson, 1966, p. 63), and "[i]f it can be registered, the organism will be able to detect the slope of the ground."

In this simple example, Gibson did several things; 1) he defined two physical referents, that is, referents that exist outside the animal as part of the physical world, 2) he established that these referents are independent—that they can vary relative to one another (e.g., the ground can be sloped, relative to the gravitoinertial force environment), 3) he pointed out that the animal can detect (and control) its orientation relative to each referent, and 4) that the simultaneous orientation of the animal relative to these referents structures emergent patterns that extend across different forms of ambient energy (gravitoinertial, and mechanical). Finally, 5) Gibson noted that these higher order patterns might be detected, as such. This may be the first presentation of a parameter of the global array.

If higher order, emergent patterns that extend across different forms of ambient energy exist in the example provided by Gibson (1966), then they can exist, in general. Thus, this example begs the questions of whether such patterns might be limited to "basic orienting" (as in Gibson, 1966), or whether their existence might be general (as we claim).

## *Exploratory activity*

Exploratory activity was central to Gibson's (1966) concept of *obtained stimulation*, with the idea that information is *available*, rather than being imposed upon us and, therefore, must be *picked up* (Neisser, 1976). In Chapters II and XII, Gibson focused on general concepts of exploratory activity; the need for it, its ubiquity, and some of its general characteristics. But examples and instances of exploratory activity appear throughout the book, as he noted (p. 250). In Gibson's analysis, information about the animal-environment interaction was generated primarily through the animal's active engagement with the environment (e.g., Heft, 1988; Richardson et al., 2008, Principle IV). This assertion has most often been qualitative (e.g., Adolph, 1995; E. J. Gibson & Pick, 2000; Gibson, 1966, 1979; Riley et al., 2002). Analytical treatments have shown that the quantitative details of information in the global array arise from the quantitative details of exploratory activity (e.g., Bingham & Stassen, 1994; Peper, Bootsma, Mestre, & Bakker, 1994). These analyses are consistent with the hypothesis that the generation and pickup of information in the global array can be (and, likely, typically is) influenced by individual differences in perceptual-motor control and skill. It is not enough that perceivers have skilled control of their own movements. In addition, perceivers must learn to select movements that generate the information that is desired, that is, exploratory movement should be selected on a task-specific basis. At a qualitative level, infants do this, deploying particular exploratory actions under some task conditions but not under others (Adolph, 1995; Adolph et al., 2000). Adults also do this, as shown by Riley et al. (2002).

Information is generated by movement in the animal-environment system. This is true of the global array, just as for any other array. Thus, the generation of information in the global array and, hence, the opportunity to detect aspects of the animal-environment system, depend upon exploratory activity. Here, we provide a quantitative example of how movement generates a parameter of the global array. Mantel, Stoffregen, Campbell and Bardy (2015) formalized a parameter of the global array specifying the definite (i.e., scaled) distance between the perceiver and an object. The distance $D$ of an object is a function of the direction $\alpha$ of the object relative to the direction of the perceiver's movement and of the rate $\dot{\alpha}$ at which this angle changes. However, it also depends on the magnitude of velocity, $V$, of the point of observation, which is not available in optics (e.g., Bingham & Stassen, 1994):

$$D = \frac{V \sin \alpha}{\dot{\alpha}} \qquad (1)$$

When the perceiver moves his/her head, the optical consequences of this motion do not occur in isolation, but are accompanied by concurrent patterns in (at least) the gravito-inertial and haptic arrays. From this perspective, Equation 1 demonstrates that "pure" optic flow is ambiguous with respect to physical dynamics. It also provides a formal description of a parameter of the global array specifying distance in an emergent relation that extends across optical ($\alpha$, $\dot{\alpha}$) and non-optical variables (*V*). Equation 1 applies to purely rectilinear movements of the point of observation. Similar formal descriptions can be obtained in case of 2D and 3D movements (Equation 2 and 3, respectively; for details see Mantel et al., 2015):

$$D = \frac{V \sin\alpha}{\dot{\theta}} \quad (2)$$

$$|D| = \frac{V|\sin\alpha|}{Q} \quad (3)$$

where $\dot{\theta}$ is the angular velocity at which the direction of the point of observation changes relative to the object (equivalently, by symmetry, $\dot{\theta}$ is the angular velocity at which the direction of the object changes relative to the point of observation), and *Q* is the norm of a rotational vector also characterizing the change in the direction of the point of observation relative to the object:

$$Q = \|\vec{\Omega}\| = \left\|\vec{i}\frac{d\vec{i}}{dt}\right\| \quad (4)$$

with $\vec{i}$ being the unit vector of a mobile base, pointing from the object toward the point of observation. Contra $\dot{\alpha}$ in the 1D model (Equation 1), in the 2D and 3D models $\dot{\theta}$ and *Q* are no longer defined relative to the direction toward which the point of observation is moving. Interestingly, Equations 1-3 simplify in case of particular movements. For example, when the movement of the point of observation is tangential to an object-centered sphere, Equation 3 becomes

$$D = \frac{V}{Q} \quad (5)$$

According to Equations 1-5, definite egocentric distance is specified in the emergent relation between optical and non-optical parameters. That is, the specificational parameter exists in the global array.

## *Detecting parameters of the global array: Empirical demonstration*

The quantitative details of movement of the embodied point of observation influence the quantitative details of the global array patterns that are generated (Mantel et al., 2015) which, in turn, influence the accuracy of perception. This influence of subtle dynamics (arising from subtle facts about the animal) implies that the movement effects discussed in an earlier section of this article are not minor variations that are irrelevant to perception; rather, that embodiment of the point of observation meaningfully affects the generation of (specifying) patterns in the global array.

In Mantel et al. (2015), seated participants wore a head-mounted display unit, through which they were shown a virtual object at eye-height. Participants were asked to judge whether or not the object was within reach. In some conditions, participants were free to move before making judgments. During these conditions, the kinematics of head movement were recorded. In the free movement condition, the display of the virtual object was driven in real time by the movement of the participant's head so as to depict a stationary virtual object at a specific distance. In a control condition, participants were instructed to remain stationary while the display of the object was driven by their own movement played back from motion recordings—in effect, they watched a movie of their own earlier activity. When the display was driven by real-time movement of the head in inertial space the global array parameter specifying definite distance was available. By contrast, this parameter was not available (did not exist), when participants were presented with an optical recording of the display being driven by their (earlier) head movements. Judgments were both more precise and more accurate when the display was driven by head movements in realtime, that is, when the relevant parameters of the global array were available. Using data on head movements Mantel et al., computed the instantaneous head velocity, the instantaneous direction of the target relative to the direction of motion, and the instantaneous velocity at which this direction changed at each moment (i.e., 100 times per second). These variables are plotted in Figure 11C. Mantel et al., then used these variables to compute the information about distance that was available in the global array, according to the parameter we formalized in Equations 1-5. This information is plotted on Figure 11B, which therefore illustrates the distance information generated by one participant during real-time head movements, according to Equations 1, 3 and 5. As can be

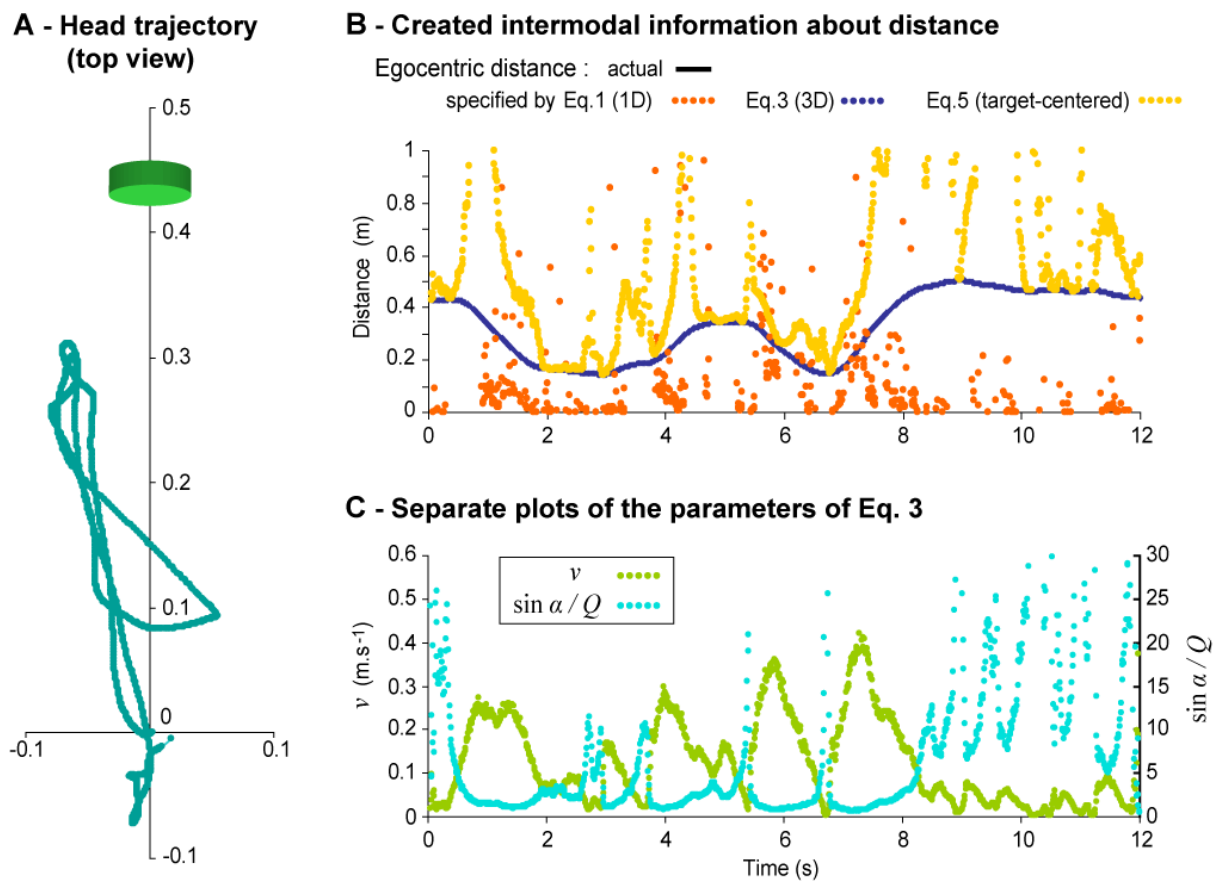

**Figure 11.** Distance information created during one representative trial in the movement condition. (A) Head trajectory (bird's eye view; in meters). The green cylinder indicates the location and size of the virtual object. (B) Evolution over time of the actual distance of the object and of the distance specified by Equations 1, 3 and 5. The curve for the actual distance is not visible on the plot because of the perfect overlap with the distance specified by Eq 3. (C) Evolution over time of the optic and inertial components of intermodal invariant as formalized in Eq 3. After Mantel et al., 2015; reproduced with permission.

27/34

seen from Figure 11C, when considered separately neither the optical variables nor the gravitoinertial variables yielded information about object distance.

The global array invariant identified by Mantel et al. (2015) does not depend upon any specific movement amplitude, velocity, or trajectory shape. However, the ability to detect the invariant may be influenced by variations in exploratory activity. For example, for a given head movement the subsequent change in direction of the object relative to the point of observation in optic flow will decrease (in magnitude, velocity, etc.) as the distance to the object increases. It follows that to maintain a minimal amount of optical movement participants would have to increase their movement amplitude, and/or velocity, acceleration and so on. Congruent with this prediction Mantel et al. (2015) found that the amplitude of head movement, instantaneous velocity and instantaneous acceleration all increased as the distance to the object increased. Mantel et al., also used Equations 1-3 and 5 to quantify, *a posteriori* the amount of accurate distance information generated by each participant's exploratory activity, that is, they computed the average amount of time that each equation specified the actual distance of the object. As expected, Equation 3 (relating to movement of the head in three dimensions) provided accurate distance information 100% of the time. In contrast, Equation 1 (related to movement of the head in one direction, only) generated accurate information only 3.5% of the time. As Equation 5 illustrates, the global array invariant undergoes a qualitative change when the direction of head motion is (at least locally) curvilinear and centered on the object of regard. Indeed, in such cases, the global array parameter specifying distance does not depend on the direction of the object relative to the direction of motion. Individual participants who more often generated accurate distance information in the global array through their movements (per equation 5) also exhibited a significantly better accuracy in their judgments of whether objects were within reach. Taken together, these results suggest that participants may have moved in such a way as to increase the availability, saliency and ease of pick-up of the relevant parameter in the global array.

## *Analytic implications*

Gibson (1966; Gibson, et al., 1955) recognized the need to formalize (i.e., to render as mathematical expressions) parameters of the optic array. He understood that formalization was required, not only to permit experimental evaluation of the perceptual reality of ambient patterns, but equally to establish the legitimacy of the concept of "ambient information" in the community of behavioral scientists. The same logic applies to understanding and analysis of the global array. Choosing to focus analytic work within a single form of ambient energy leaves such analyses open to "brittle" falsification via existence proofs, but also prohibits access to (or testing of) patterns that exist in the global array. Stoffregen and Bardy (2001) offered examples of parameters of the global array that had been formalized. Mantel et al. (2015) derived a new formalization and used it to make and test predictions about perception under experimental conditions.

## *Methodological implications*

In general, future research about the global array should formalize new task-specific invariants in the global array, test perceptual sensitivity to these invariants, and evaluate experimentally how these invariants are used in perception and action situations. Researchers must design experimental manipulations that make it possible to vary independently parameters in single-energy arrays versus parameters in the global array. Only with such independent control is it possible to evaluate the relative contribution of each to perception and action (cf. Gibson, 1966,

p. 63). Stoffregen and Bardy (2001, Section 7) described how this could be done. Mantel et al. (2015) used this program to develop and implement an experimental setting in which parameters of the global array could be manipulated independently of parameters in component single-energy arrays. Their study demonstrated that this type of experimental method was possible, and that the use of such methods could yield qualitatively new results.

## Conclusion

In 1966, Gibson introduced the concept of the perceptual system. This concept was based upon the obtaining of ambient information by active perceivers. In *The Senses Considered as Perceptual Systems*, Gibson rejected ideas that had been central to perceptual scholarship for many centuries. However, his rejection was not complete. In the present article, we have argued that Gibson did not reject one ancient assumption—that perception is divided into distinct, divisible types. Gibson rejected classical ideas of the senses as passive channels of sensation, but he accepted the plurality of the senses, and proposed a plurality of perceptual systems. Stoffregen and Bardy (2001) introduced the concept of the global array. They argued that information sufficient for accurate perception (and, therefore, enabling direct perception) exists in the global array, and that such information exists exclusively in the global array. In the present article we have argued that, taken to its logical conclusion, Gibson's concept of lawful, 1:1 specification in ambient energy patterns, together with his concept of the perceptual system, leads inexorably to two conclusions. First, as argued by Stoffregen and Bardy, information in the Ecological sense exists exclusively in the global array, and second, in obtaining information from the global array our sensory apparatus functions as a single perceptual system, unitary and indivisible.

One of the defining features of Gibson's career was his willingness to hold up for explicit scrutiny ideas that, for hundreds or even thousands of years, had been accepted as being uncontroversial, or even self-evident. In 2017, it can be difficult to appreciate how startling, even bizarre many of Gibson's claims were for his contemporaries. But the extraordinary nature of Gibson's ideas was apparent to many. For example, Hochberg (1971, p. 505, emphasis added), acknowledged that "the potential importance of [Gibson's arguments] can hardly be overstated. If Gibson is right, we must replace *all previous assumptions* about what are the adequate units of stimulation to which we respond…". Similarly, Sherrick (1967, p. 529) suggested that Gibson had "shaken the reader's faith in the orthodoxy of perceptual theory from Aristotle to Boring." We believe that this same willingness to scrutinize and reject ancient assumptions should be applied to what we have called the assumption of separate senses. If the assumption of separate senses is rejected, then it becomes possible to question and reject the assumption that lawful, 1:1 specification can exist in patterns existing in individual forms of ambient energy. We see this as a great challenge for the Ecological Approach to Perception and Action: Within the Ecological approach, is it possible to retain both the assumption of separate senses and the claim that the animal-environment interaction is specified in patterns of ambient energy?

Stoffregen and Bardy (2001) presented the global array in a target article in *Behavioral and Brain Sciences*. The target article appeared in print with many commentaries from scholars in diverse fields, including adherents of the Ecological Approach to Perception and Action. Some of these commentaries were supportive (e.g., Adolph, Marin, & Fraissse, 2001; Walker-Andrews, 2001), while others raised objections (e.g., Flom & Bahrick, 2001). In most cases, objections took the form of examples of situations that, in the authors' view, constituted exceptions to our claim that specification exists exclusively in the global array. In responding to the commentaries,

Stoffregen and Bardy (2001) addressed these individual cases and, informally, we have continued to do so in subsequent discussions. Instances are important, but they are not more important than general arguments. Our theory of the global array is supported by examples (instances) but it is based on general principles. Accordingly, our argument cannot be rejected (or even fairly evaluated) solely on the basis of examples, or instances. We hope that the present article (together with Stoffregen & Bardy, 2001) will motivate adherents of the Ecological Approach to Perception and Action to begin to address the relevant issues not only at the level of instances, but also at the level of general principles. As one example, any claim that some facts are specified in the global array while others are specified in a single-energy array ultimately requires some *a priori* argument establishing a boundary between "things that are specified in the global array", and "things that are specified in single-energy arrays". Any such boundary should be law-based, and cannot be justified solely on the basis of instances. As another example, the assumption of separate systems is an assumption, and as such it requires *a priori* arguments, both in logical terms and in terms of biological evolution. As Stoffregen and Bardy (2001, Section R8) pointed out, instances that are claimed to exemplify specification in single-energy arrays imply the assumption (typically, implicit) of the existence of separate senses.

Perhaps the simplest (and therefore, the most easily neglected) lesson of *The Senses Considered as Perceptual Systems* is that perception is multisensory: The animal-environment interaction naturally alters the structure of many forms of ambient energy, and when an animal is awake and alert, all sensory receptors are "working" all of the time. We live in a world that is inescapably multisensory. Scientists typically focus on individual senses (or perceptual systems). We argue that this practice, whether analytical (i.e., in attempting for formalize specifying parameters) or experimental (i.e., in conducting research to identify causality in perception) is a form of reductionism that is not motivated by the ecological physics of the animal-environment system. We began by quoting Gibson's understanding that his presentation was a beginning rather than an end; the first statement of a new theory that could be completed and rendered fully coherent and consistent only through future development: "The answers to these questions are not yet clear, but I am suggesting new directions in which we may look for them", (p. 5). We believe that we are taking Gibson's (1966) ideas to their logical conclusion.

## Acknowledgements

We thank Pablo Covarrubias, Felipe Cabrera, Ángel Jiménez, and Alan Costall for their patience and forbearance in the creation of this article. Thomas A. Stoffregen offers his thanks to Trish Walsh for her love and support—while they lasted.